\documentclass[aps,twocolumn,showpacs,showkeys,preprintnumbers,amsmath,amssymb]{revtex4}

\usepackage{epsfig}

\def\be{\begin{equation}}

\def\ee{\end{equation}}

\def\lsim{\raise0.3ex\hbox{$<$\kern-0.75em\raise-1.1ex\hbox{$\sim$}}}

\def\gsim{\raise0.3ex\hbox{$>$\kern-0.75em\raise-1.1ex\hbox{$\sim$}}}

\begin{document}

\title{Damage Spreading and Opinion Dynamics on Scale Free Networks}

\vskip 1.cm

\author{Santo Fortunato}

\vskip0.7cm

\affiliation{Fakult\"at f\"ur Physik, Universit\"at Bielefeld, D-33501 Bielefeld, Germany}

\vskip0.5cm

\begin{abstract}
\noindent

We study damage spreading among the opinions of a system of agents, subjected to
the dynamics of the Krause-Hegselmann consensus model. The damage
consists in a sharp change of the opinion of one or more agents in the initial random
opinion configuration, supposedly due to some external factors and/or events. 
This may help to understand for instance
under which conditions
special shocking events or targeted propaganda are able to influence the 
results of elections. 
For agents lying on the nodes of a Barab\'asi-Albert 
network, there is a damage spreading transition at a low value 
$\epsilon_d$ of the confidence
bound parameter. Interestingly, we find as well that there is some critical
value $\epsilon_s$ above which the initial perturbation manages to propagate
to all other agents.

\end{abstract}

\pacs{87.23.Ge, 89.75.Hc}

\keywords{Networks, damage spreading, sociophysics}

\maketitle

\vskip0.7cm

Most natural, social and technological systems are continuously subjected to
external stimulations of all kinds. Examples are infections, 
hacker or terrorism attacks, noise, errors, etc. 
It may happen that, due to these shocks,
a sudden change occurs 
in some feature of a limited number 
of subjects (thinking about human beings the feature could be health, 
religion or, like in this paper, opinion),
and that successively, through interactions with other 
members of the system, this perturbation spreads 
until it eventually affects a big portion of the system.
The study of such processes is of great importance, in order to
take the effects of eventual future local failures under control, but it 
is also an important method to investigate the dynamics of a system.   

Damage spreading (DS) was originally used by Kauffman \cite{kauf} as a tool for studying
biologically motivated dynamical systems. 
In physics, the first investigations focused on 
the Ising model \cite{stasta}. Here one starts 
from some arbitrary configuration of spins and creates a replica by flipping 
one or more spins; after that one lets both configurations evolve towards equilibrium
according to the chosen dynamics 
under the same thermal noise (i.e. identical sequences of random numbers).
It turns out that there is a temperature
$T_d$, near the Curie point, which separates a phase where the
damage heals from a phase in which the perturbation 
extends to a finite fraction of the spins of the system. 

Meanwhile
there is a sizeable literature on this problem,
which finds applications in many fields of modern science.
Epidemiology, for example, is by definition the study of a particular
DS problem \cite{and}. After the 
recent discovery that 
many systems in nature and society
can be described as complex networks,
scale free graphs have been intensively 
investigated and many classical problems have been reformulated
on such special topologies
\cite{netw}. In particular it is very interesting to understand
the mechanisms by which diseases, information, computer viruses, etc. spread
over networks. 

In this paper the network represents the system of acquaintances
between people and
we study the following problem: suppose we have a 
community of voters at the beginning of an electoral campaign, 
during which the voters shape their own opinions through relationships
with their friends.
If a small set of voters for any reason suddenly change their  
mind at this initial stage, 
would the final outcome of the election be influenced and,
if yes, to which extent? 
Very recent history delivers a dramatic example:
the shock caused to the Spanish people by the 
bombs in Madrid
on March the 11th 2004 turned over the outcome of the national elections,
which seemed already decided till that day. 

What we need is a model that describes how people
convince each other. 
In the last years quite a few models of opinion dynamics have been proposed, 
like those of Galam \cite{Galam}, Deffuant et
al. \cite{Deff}, Krause-Hegselmann (KH) \cite{HK}, or Sznajd \cite{Sznajd}, and 
sociophysics simulations have become a fruitful field of research
\cite{weidlich, staufrev}.
As far as our DS problem is concerned, some results on the Sznajd model recently
appeared \cite{sorn}, but for an improbable society where agents sit on 
the sites of a square lattice.
 
Here we present the first systematic study on the subject. Our agents are on the
nodes of a Barab{\'a}si-Albert (BA) network \cite{BA}, which represents a more
realistic model for the structure of social relationships. 
The network is
constructed by means of a growth process starting from $m$ nodes which are all
connected to each other. Nodes are then added one by one and each of them forms $m$ edges 
with the existing vertices, such that 
the probability to get linked to a node is proportional
to the number of its neighbours. At the end, 
the number of agents with {\it degree} $k$,
i.e. having $k$ neighbours, is proportional to $1/k^3$ for $k$ large,
independently of $m$. 
Our simulations show that the main results are only weakly dependent on $m$, so we
focused on the case $m=3$. 

We adopted the opinion dynamics of the KH consensus model. In this model,
the opinions are real numbers between $0$ and $1$
and a confidence
bound parameter $\epsilon$, also real in $[0:1]$, is introduced. One starts from a random
distribution of opinions.
If we want to update the opinion $s_i$ of agent $i$ we have to select among all
neighbours of $i$ only those agents whose opinions are compatible with $s_i$, i.e. 
those agents $j$ such that $|s_i-s_j|<\epsilon$; next, $i$
takes the average opinion of the compatible agents.
We chose to update sequentially the status of the agents, 
in an ordered sweep over the whole population; in this way, the dynamics does
not require random numbers and is therefore truly deterministic, 
at variance with other consensus models.
At some stage, the system will converge to a configuration 
that the dynamics is unable to 
modify. 
This configuration represents the equilibrium state and
the final opinion distribution, which is given by a set of $\delta$-functions, 
depends on 
$\epsilon$.
On BA networks, we found that the threshold for complete consensus
is $\epsilon_c=1/2$, independently of $m$. 

For the DS analysis we followed the procedure that we exposed
above for the Ising model. 
After creating a random opinion configuration, we produced a replica of
it in that we changed the opinion of a single agent $i$. 
The results do not
depend on the exact number of perturbed sites, as long as they 
are a fraction of the population $N$ 
that vanishes in the limit 
when $N$ goes to infinity \cite{foot2}. 
We perturbed the opinion as follows:
if $s_i>1/2$ the new
opinion becomes $s_i-1/2$, otherwise $s_i+1/2$. 

\begin{figure}[htb]
  \begin{center}
    \epsfig{file=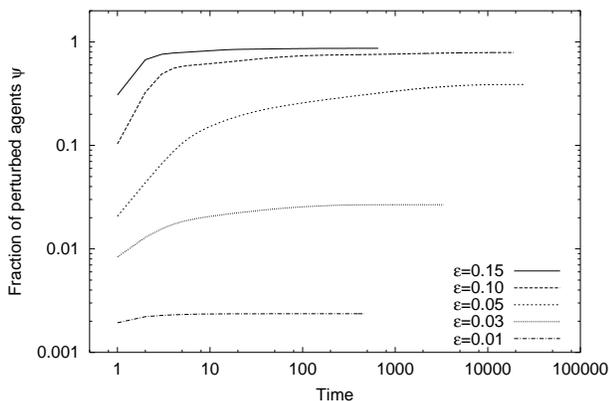,width=9.0cm}
    \caption{\label{fig1}{Time evolution of the fraction of perturbed opinions
        for several values of $\epsilon$. Here the population is $N=10000$.}}
  \end{center}
\end{figure}

After each sweep over the network we calculated 
the fraction $\Psi$ of different opinions in the two configurations.
The simulation stops when both systems reach their final states, which 
happens when no agent changed opinion during an iteration.
In all our calculations we used double 
precision real numbers and we decided that two opinions are the same if 
they differ from each other by less than $10^{-9}$, otherwise they are different. 
For the results to have statistical significance we took averages over $1000$ samples
for all values of $\epsilon$ and of the population $N$.

We stress that on our network, like in a real society, some nodes are
more important than others. If we aim at spreading an opinion in the society,
we should better try to convince people with many friends 
than persons with few social contacts. We then expect that 
the perturbation on the whole system will be more relevant
if we initially damage a hub than a loosely connected node and we investigated
both situations. 

First, we analyzed the case in which the perturbed node is a hub.
Fig. \ref{fig1} shows the variation with time of $\Psi$ (the time unit is 
one sweep over the network). There is a characteristic
pattern with an initial phase in which $\Psi$ grows rapidly, followed by
a very slow relaxation to the final value.  

\begin{figure}[htb]
  \begin{center}
    \epsfig{file=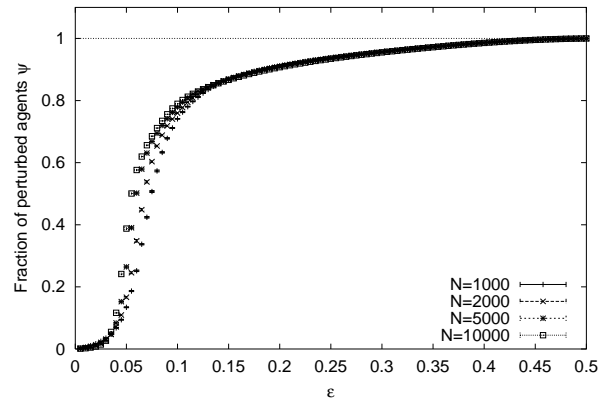,width=9.0cm}
    \caption{\label{fig2}{Variation of $\Psi$ with the confidence bound
        parameter $\epsilon$.}}
  \end{center}
\end{figure}

As we can see from the figure,
the relaxation time increases up to $\epsilon\,\sim\,0.1$, then it decreases.
At the end of relaxation the number of damaged sites stays constant
although the system keeps evolving. 
This is due to the fact that, at advanced stages of the evolution, the 
opinions are grouped in clusters.
Agents of different clusters cannot interact with each other,
as their opinions differ by more than $\epsilon$. Because of that, if a cluster
contains perturbed opinions, most of its agents will be sooner or later affected
\cite{foot3}, 
otherwise it can never be reached by the perturbation. 

\begin{figure}[htb]
  \begin{center}
    \epsfig{file=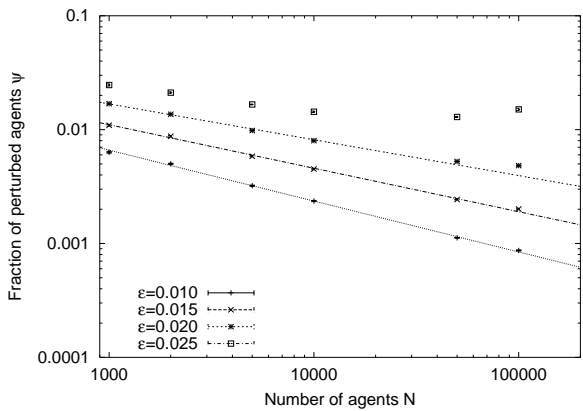,width=9.0cm}
    \caption{\label{fig3}{Dependence of $\Psi$ on the population $N$
        for low values of $\epsilon$.}}
  \end{center}
\end{figure}

In Fig. \ref{fig2} we plot $\Psi$ as a function of $\epsilon$, for 
different network sizes.
We see that the damage rises fast with $\epsilon$ and that, for 
$\epsilon$ larger than about $0.05$, more than half of the agents have been affected.
The inflection of the curves at low values of the confidence bound 
hints to the existence
of a DS transition, like in the Ising model. 
If there is indeed a phase in which the damage affects only a vanishing fraction
of the system, we should find that there $\Psi$ goes to zero when 
$N$ goes to infinity.
We then looked for scaling behaviour of $\Psi$ with $N$. In Fig. \ref{fig3}
we plot $\Psi$ as a function of $N$ for several values of the confidence bound.
We can see that the points can be quite well fitted by a simple power law up to
$\epsilon\sim\,0.015$. For higher values, a saturation to a non-zero $\Psi$ takes place.
As estimate of the DS threshold $\epsilon_d$, we took the value which gave
the smallest $\chi^2$ for the fit with the power law: we found
$\epsilon_d=0.013(3)$. The error marks the range where the 
above-mentioned $\chi^2$ (per degree of freedom) is below $1$.

Coming back to Fig. \ref{fig2}, we notice that in contrast to
the Ising or Kauffman models, we have a second
threshold $\epsilon_s$ above which all agents are affected by the initial local
perturbation. To our knowledge, this is a truly novel feature for a DS process,
and we expect it to hold as well for the opinion dynamics of Deffuant et al.
By studying the dependence of this threshold on $N$, we extrapolated 
its infinite-$N$ limit $\epsilon_s=0.500(1)$. We remark that this is  
just the threshold for complete consensus of our system. So, if we are in the 
consensus regime and a single agent suddenly changes its mind, this suffices
to (slightly) modify the final dominant opinion of the total population.


We tried to check whether this unexpected feature is specific of
the particular social topology we have chosen, or whether it is 
exclusively due to the dynamics. 
Simulations of a society where agents sit on the sites of a square lattice,
with periodic boundary conditions, confirm the result and $\epsilon_s$ is
again $1/2$.
We have examined as well a community where each agent has
relationships with all others; here we have that $\epsilon_s$ 
is about $0.07$, much lower than the consensus threshold $0.21$.
For this special
society there is no DS transition, as each agent effectively 
interacts with a finite
fraction of the system, and it can be proved that $\epsilon_d(N)$
vanishes when $N\rightarrow\infty$.

We remind that we have introduced the damage right at the beginning of the
evolution. We have also performed some tests to check what happens
if we instead perturb the system after some evolution steps.
Now the 
perturbed agent stays in a society where people are mostly divided
in groups of close opinions, and such communities will 
evolve separately from each other. In this way the 
shocked agent can interact only with a smaller portion 
of the system, i.e. with a few clusters of agents,
and the amount of damage will drop. On the other hand, we find that both $\epsilon_d$ and
$\epsilon_s$ remain the same.

\begin{figure}[htb]
  \begin{center}
    \epsfig{file=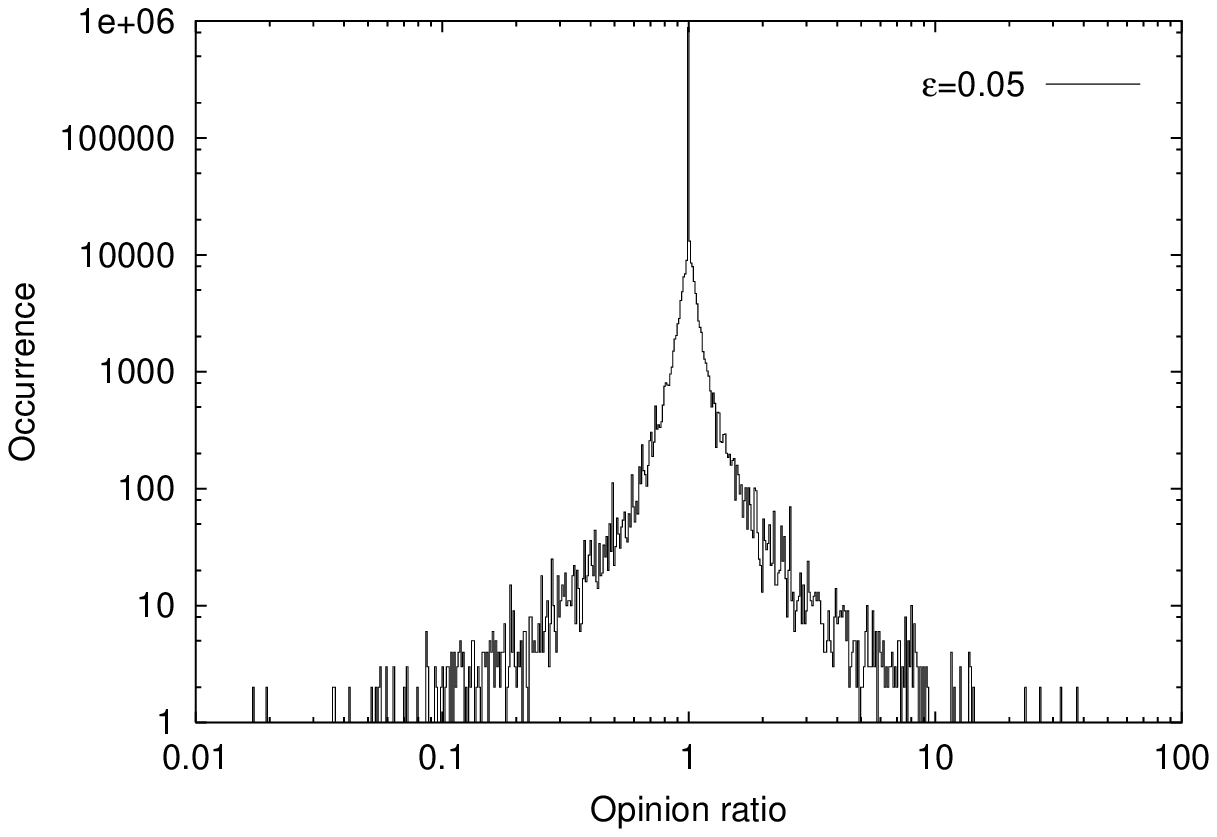,width=9.0cm}
    \epsfig{file=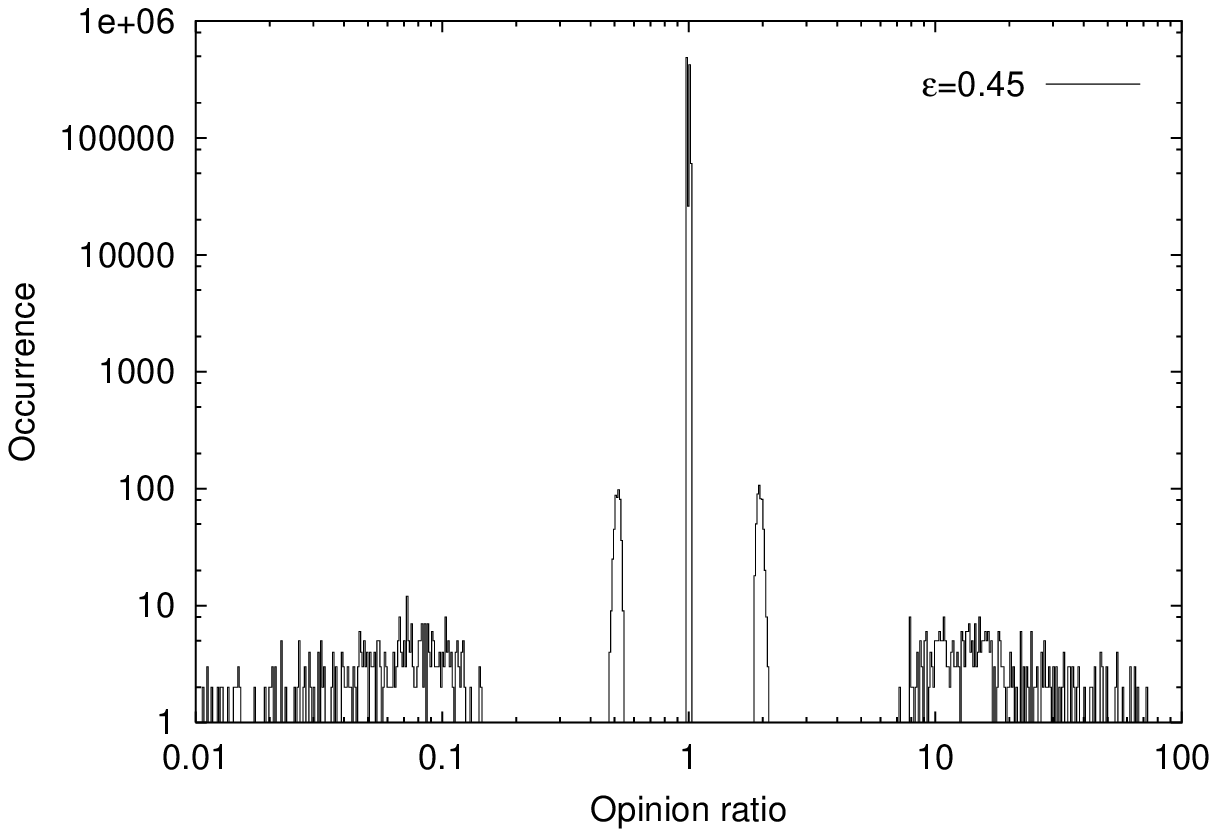,width=9.0cm}
    \caption{\label{fig6}{Histogram of the ratios of the agents' opinions
with and without damage. The confidence bound 
is $\epsilon=0.05$ (top) and $0.45$ (bottom); the population is $N=1000$.}}
  \end{center}
\end{figure}

Furthermore we have studied the effective opinion variations of the agents 
if damage is introduced.
For this purpose we divided the opinion of each agent in the 
perturbed configuration by the corresponding value in the unperturbed 
configuration.
In Fig. \ref{fig6} we show the distribution of the 
opinion ratios at the end of the time evolution, 
for two values of the confidence bound.
As we can see, when $\epsilon$ is small, the distribution 
is strongly peaked at the value $1$, and the opinions vary continuously though 
very little in most cases.
As we approach the consensus threshold, instead, we notice that the opinion variations
are no longer continuously distributed, and other narrow peaks appear, which 
shows that most values of the ratios are suppressed and discontinuous 
jumps, corresponding to drastic opinion changes, are allowed.

\begin{figure}[htb]
  \begin{center}
    \epsfig{file=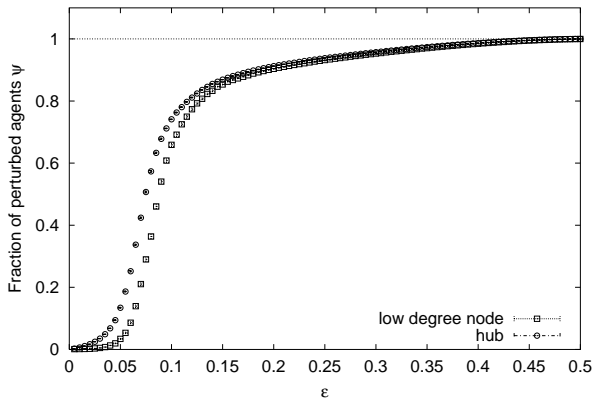,width=9.0cm}
    \caption{\label{fig5}{Comparison of the DS curves corresponding to the
initial perturbation of a node with low degree (squares) or a hub (circles).
The number of agents is $N=1000$.}}
  \end{center}
\end{figure}

Let us now check what happens when the initially damaged agent 
sits on a node with low degree. In Fig. \ref{fig5} we 
compare the DS curve obtained in this case with the 
curve for a perturbed hub.  
We see that, for any $\epsilon$, $\Psi$ is larger
when we damage the hub, as expected.  
On the other hand
we find that both $\epsilon_d$ and $\epsilon_s$ are the same as before.
That relies on the small world effect \cite{watts}
on scale free networks like ours.
In fact, each node can reach any other through a small number 
of intermediaries. In this way, even if we perturb a loosely connected node,
within the first evolution steps the perturbation will have reached quite a few
nodes with much higher degrees, which brings us back to the previous case.
The damage is larger if we perturb a hub because more agents can be reached at
the beginning of the evolution; soon after that, as we said above, clusters of opinions
are formed which do not interact with each other and the perturbation
can exclusively spread within the affected clusters. 
In conclusion, due to the small world effect, 
$\Psi$ and the fraction $\Phi=1-\Psi$ of unperturbed agents
are of the same order of magnitude in both situations:
if $\Psi$ ($\Phi$) is zero in one case, it will be zero in the other case too.

We have studied damage spreading for the Krause-Hegselmann opinion dynamics on 
Barab\'asi-Albert networks.
We distinguish three phases in the
confidence bound space, corresponding to zero, partial and total damage, respectively.
The existence of a phase where the perturbation affects all agents
is new for damage spreading processes and is
independent of the social topology, so it only relies
on the dynamics. This feature seems unrealistic, but we should 
consider that our dynamics is very simple \cite{braz}.
Moreover, we let 
our system reach in any case its final state, but the evolution time 
grows with the size of the community and would be very long for a 
realistic population of voters; normal electoral campaigns last a few months, so
we should interrupt the evolution process at an earlier stage and the damage
could then be limited.
The probability to have large 
variations of the final agents' opinions induced by the
initial perturbation is rather small, 
but it increases with $\epsilon$ and,
by approaching the consensus threshold, forbidden bands appear. 
The amount of the damage depends on the degree of the damaged node but
the thresholds for damage spreading and    
saturation do not, because of the small world effect.

In the future it would be helpful to use other opinion dynamics
to check for the consistency of the results. 
For more realistic analyses of the problem, together with 
eventual refinements of the existing consensus models and of the
social topology, 
it is important to include as well other factors in the 
dynamics, like advertising and noise.

I gratefully acknowledge the financial support of
the TMR network ERBFMRX-CT-970122 and the DFG Forschergruppe FOR
339/1-2.


\begin{thebibliography}{99}

\bibitem{kauf} S. A. Kauffman, J. Theor. Biol. {\bf 22}, 437 (1969).

\bibitem{stasta} M. Creutz, Ann. Phys. {\bf 167}, 62 (1986);
H. E. Stanley, D. Stauffer, J. Kert{\'e}sz and H. J. Herrmann,
  Phys. Rev. Lett. {\bf 59} 2326 (1987).

\bibitem{grass1} P. Grassberger, J. Phys. A {\bf 28}, L67 (1995).




\bibitem{and} R. M. Anderson and R. M. May, {\it Infectious Diseases of Humans},
Oxford University Press, Oxford (1991); N. T. J. Bailey, {\it The Mathematical Theory
of Infectious Diseases and its Applications}, Hafner Press, New York (1975).

\bibitem{netw} R. Albert and A. L. Barab{\'a}si, Rev. Mod. Phys. {\bf 74}, 47 (2002);
M. E. J. Newman, SIAM Review 45, 167-256 (2003).

\bibitem{Galam} S. Galam, J. Stat. Phys. {\bf 61}, 943 (1990) and Physica A 
  {\bf 238}, 66 (1997).

\bibitem{Deff} G. Deffuant, D. Neau, F. Amblard and G. Weisbuch,
Adv. Complex Syst. {\bf 3}, 87 (2000); G. Weisbuch, G. Deffuant, F. Amblard, and
J.-P. Nadal, Complexity {\bf 7}, 2002; G. Deffuant, F. Amblard, G. Weisbuch and 
T. Faure, Journal of Artificial Societies and Social Simulations {\bf 5}, issue
4, paper 1 (jasss.soc.surrey.ac.uk) (2002).

\bibitem{HK} R. Hegselmann and U. Krause, Journal of Artificial Societies and 
Social Simulation {\bf 5}, issue 3, paper 2 (jasss.soc.surrey.ac.uk) (2002) and
Physics A, in press (2004); U. Krause, {\it Soziale Dynamiken 
mit vielen interakteuren. Eine Problemskizze}.
In U. Krause and
M. St{\"o}ckler (Eds.), {\it Modellierung und
Simulation von Dynamiken mit vielen interagierenden Akteuren}, 37-51,
Bremen University, Jan. 1997.

\bibitem{Sznajd} K. Sznajd-Weron and J. Sznajd, Int. J. Mod. Phys. C {\bf 11},
  1157 (2000).

\bibitem{weidlich} W. Weidlich, {\it Sociodynamics; A Systematic Approach to 
Mathematical Modelling in the Social Sciences}. Harwood Academic Publishers,
2000.

\bibitem{staufrev} D. Stauffer, {\it The Monte Carlo Method on the Physical Sciences},
  edited by J. E. Gubernatis, 
AIP Conf. Proc. {\bf 690}, 147 (2003), cond-mat/0307133.

\bibitem{sorn}B. M. Roehner, D. Sornette and J. V. Andersen, {\it Response
    Functions to Critical Shocks in Social Sciences: An Empirical and Numerical
    Study}, in press for Int. J. Mod. Phys. C, issue 6 (2004).

\bibitem{BA} A. L. Barab{\'a}si and R. Albert, Science {\bf 286}, 509 (1999).


\bibitem{foot2} Indeed, some test runs we performed with an 
initial perturbation of as many as ten 
agents led to the same results as with a single 
damaged agent, within errors.

\bibitem{foot3} It may happen that the neighbours of an agent
are incompatible with it: in this case the opinion
of the agent cannot change, so it cannot be perturbed either. 

\bibitem{watts} D. J. Watts, S. H. Strogatz, Nature {\bf 393}, 440 (1998).

\bibitem{braz} Nevertheless, with a simple model of this kind 
(that of Sznajd) one could reproduce 
the results of Brazilian and Indian elections, see 
A. T. Bernardes, D. Stauffer and J. Kert{\'e}sz, 
Eur. Phys. J. B {\bf 25}, 123 (2002) and 
M. C. Gonzalez, A. O. Sousa and H. J. Herrmann, 
Int. J. Mod. Phys. C {\bf 15}, No. 1 (2004).

\end{thebibliography}
\end{document}